# Refnd: Preventing Data Leakage in Relational Datasets

**Anthony Lavertu**[1] **Jacob Côté**[1] **Jacques Corbeil**[2 3]
**Sophie Gobeil**[4] **Pascal Germain**[1]

[1]Department of Computer Science and Software Engineering, Université Laval, Québec, Canada
[2]Department of Molecular Medicine, Université Laval, Québec, Canada
[3]Mila - Quebec Artificial Intelligence Institute, Montreal, Canada
[4]Département de biochimie, de microbiologie et de bio-informatique, Université Laval, Québec, Canada

anthony.lavertu.1@ulaval.ca jacob.cote.3@ulaval.ca
jacques.corbeil@fmed.ulaval.ca
sophie.gobeil@bcm.ulaval.ca Pascal.Germain@ift.ulaval.ca

## Abstract

Machine learning models trained on biochemical data are routinely evaluated using splits that fail to account for relational structure, causing information leakage and over-optimistic performance estimates. Existing splitting methods lack theoretical grounding and scale at best quadratically. We introduce the Relational Generative Process (RGP), a mathematical formalization explaining why relational structure arises in biochemical datasets, and Refnd, a splitting algorithm that leverages a proximity graph computed in loglinear time using Hierarchical Navigable Small World (HNSW). We validate on an antimicrobial peptide dataset, showing that Refnd splits yield lower but more realistic evaluation performance than traditional splits. Refnd is applicable to any dataset arising from an RGP such as protein sequences and structures, small molecules, and nucleotide sequences, and is openly available as a Rust accelerated Python package: `pip install refnd`.

## 1 Introduction

Reliable evaluation of machine learning models is a prerequisite for safe and trustworthy deployment as overestimated generalization performance yields models that appear strong under evaluation, but fail to reach those performances on new data [1]. The standard approach is to randomly partition the dataset into two subsets: training and test, where the test is held out from training to estimate real-world performance [2]. This works when the core assumption of independent datapoints is respected. However, datasets frequently exhibit relational structure, where pairs of samples share some form of relation, violating the independent assumption [3]. When such relations exist, random splitting no longer guarantees an independent hold-out set, as the resulting train and test sets will systematically share dependent samples [3]. This is especially problematic when relations correlate with the problem target: the test set can no longer be considered unseen as a model can exploit these relations to predict the target without learning the underlying mapping, effectively using information at evaluation time that would not be available at inference. This is the textbook definition of data leakage, and results in an overoptimistic and therefore misleading evaluation performance [4] [5] [6].

This failure case becomes particularly insidious when inter-sample relations are latent and must be inferred from the data. We formalize a generative process that we refer to as the **Relational Generative Process (RGP)**, that explains the emergence of a relational structure between samples in a dataset and how it can be inferred. This formalization is what fundamentally distinguishes our approach from prior work: existing methods construct splits through empirical heuristics, without theoretical grounding for why a particular partitioning strategy is the correct operation [7] [8] [9]. The RGP provides exactly this justification, and from it we rationally derive Refnd, a principled splitting algorithm with $O(n \log(n))$ complexity — in contrast to the $O(n^2)$ complexity of existing approaches [7] [9].

The RGP framework finds broad instantiation in biochemical data. Proteins, peptides, and genes acquire relational structure through evolutionary incremental modification [10]: sequences descended from a common ancestor tend to be similar, and similar sequences tend to share biological function [11]. An analogous process governs pharmaceutical small molecule datasets, where the discovery of an active scaffold is typically followed by systematic structural variations to fine-tune its activity [12] [13]. In both cases, related samples share structural similarity and target properties — the condition under which random splitting induces leakage [3] [14].

We leverage the relational structure prevalent in biochemical datasets to empirically validate Refnd. Specifically, we select an antimicrobial peptide potency prediction task as a particularly stringent test case, given the dense inter-sample relational structure of the dataset [15] [16]. Our results show that Refnd yields lower but more conservative evaluation performance than both random and a domain-specific similarity-based split, consistent with a meaningful reduction in information leakage and potentially a more faithful estimate of real-world generalization. Our contributions are threefold: (1) a mathematical formalization of the RGP and its implications for dataset splitting, (2) a principled splitting algorithm with $O(n \log(n))$ complexity, and (3) empirical validation demonstrating the practical impact of split quality on reported model performance.

## 2 Related Work

Random splitting is a common data partitioning strategy across machine learning, yet in biochemical applications it regularly yields over-optimistic evaluation performance by failing to account for relations in the data [15]. For protein sequences specifically, database clustering tools such as CD-HIT [7] and MMSeqs2 [8] have been adopted as a more principled alternative. These tools cluster sequences by similarity, and datasets are then split along cluster boundaries such that no cluster spans both train and test sets. However, both tools rely on heuristics to reduce computational cost, introducing strong assumptions about the data that do not always hold for a strict dataset split. As a result, highly similar sequences can end up in different clusters, reducing but not eliminating information leakage compared to random splitting, as we empirically demonstrate in our results.

Recent years have seen a wave of methods that address this limitation more rigorously by grounding the split in a proximity graph, where samples are nodes and edges connect pairs whose similarity exceeds a threshold. Unlike CD-HIT and MMSeqs2, which are restricted to protein and nucleotide sequences, these approaches are applicable to a broader range of biochemical data types, including small molecules and protein structures. These methods differ in how they partition the graph into training and test sets. On the more computationally intensive end, DataSAIL [17] formulates the partition as a combinatorial optimization problem, and Lo-Hi [18] frames it as a Balanced Vertex Minimum $k$-Cut problem. Other approaches adopt simpler partitioning strategies: GraphPart [19] uses restricted single-linkage clustering; Hestia [9] and AutoPeptideML [20] split along connected components; SpanSeq [21] uses DBSCAN [22] to identify clusters; and QMAP [15] detects dense communities via the Leiden algorithm [23]. Despite their differences, these approaches share two fundamental limitations. First, their partitioning strategies are empirically motivated rather than theoretically grounded, which makes their use look *ad hoc*. Second, they all operate on a pairwise similarity matrix, resulting in at least $O(n^2)$ complexity. Attempts to mitigate this use data-specific filters that skip redundant or irrelevant similarity computations, which reduce runtime, but rely on strong assumptions that frequently fail on diverse datasets, without improving the asymptotic complexity.

In this work, we address these limitations by introducing **Re**lation **Find**er (Refnd). We first provide a mathematical formalization — the Relational Generative Process — that motivates and justifies graph-based dataset partitioning from first principles, justifying the empirically motivated, *ad hoc* choices of prior work. We then bypass pairwise matrix computation entirely by directly approximating the proximity graph using the Hierarchical Navigable Small World (HNSW) algorithm [24], achieving $O(n\log(n))$ complexity under minimal assumptions, making Refnd both more scalable and more robust than existing approaches.

# 3 Methods

## 3.1 Problem Formulation

Let $\mathcal{D}$ denote the data-distribution. We consider that the data-generating process leading to $\mathcal{D}$ is structured into two steps. First, a graph structure $g$ is sampled i.i.d. from a meta-distribution $\mathcal{G}$. Second, an observed sample $x$ is a node of the graph $g$. We consider that each graph comes with its own distribution such that

$$x \sim \mathcal{D}_{|g}, \quad \text{with } g \sim \mathcal{G}, \tag{1}$$

In practice, a dataset contains multiple observations from the same graph structure distribution. Thus, we model the dataset generation process by first sampling $m$ i.i.d. graph structures, that is

$$g_i \sim \mathcal{G}, \quad \text{for} \quad i = 1, ..., m, \tag{2}$$

and then sampling $N_i$ observations for each structure

$$S = \cup_{i=1}^{m} S_i \quad \text{with} \quad S_i = \left\{x_{i1}, ..., x_{iN_i}\right\} \sim \mathcal{D}_{|g_i}^{N_i}, \tag{3}$$

The resulting dataset structure of such generating process is illustrated in Figure 1. We refer to this generative process as a Relational Generative Process (RGP) because of the relations that appears between some samples. It can happen that the graph structure $g$ is latent, in such a case the relation between the sample becomes unknown. This formulation – with a latent graph structure $g$ – broadly applies to biochemical data [9]. Proteins, peptides, and genes acquire relational structure through evolutionary incremental modification [10]. The evolution path and all intermediate instances along that path can be viewed as the latent graph structure. An analogous process governs pharmaceutical small molecule datasets, where the discovery of an active scaffold is typically followed by systematic structural variations to fine-tune its activity [12] [13], which also result in a latent graph structure.

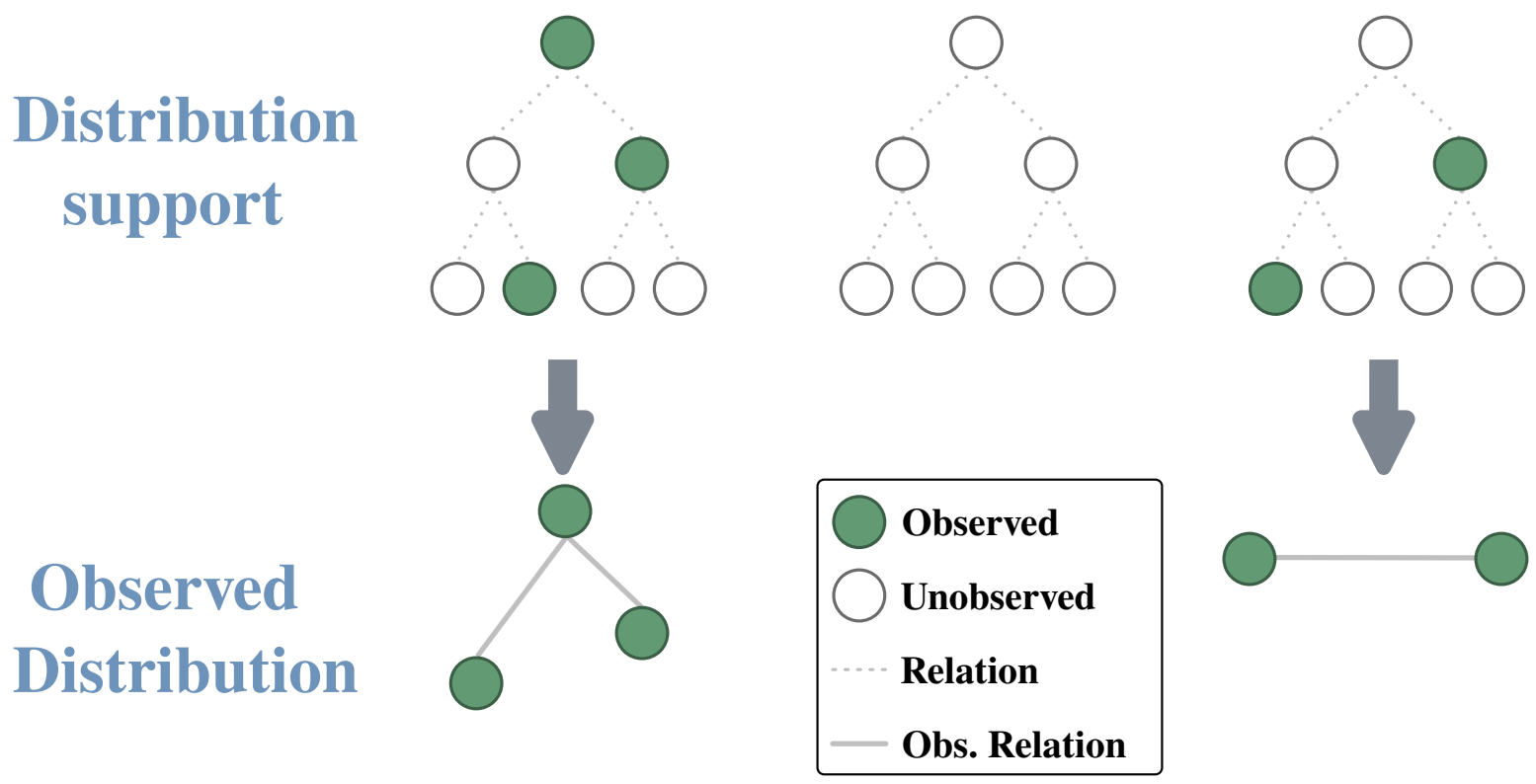

Figure 1: Illustration of the Relational Generative Process (RGP) and the resulting structure in the observed dataset. **Top**: For this illustration, we consider a finite distribution support composed of three graph structure and 7 total samples per graphs. Dotted lines represent the latent relations, green nodes represent samples observed in the dataset, and white nodes represent unobserved samples. As the full distribution support is never observed, some graph structure may contribute no samples to the dataset, as illustrated by the middle tree. **Bottom**: The observed dataset and its inferred relational structure. Solid edges represent relations inferred from sample distance, connecting samples who are close enough, suggesting they originate from a common graph. Edge length illustrate the measured distance between observed samples.

In many RGPs, samples near each other in a graph structure $g_i$ tend to exhibit similar properties, which creates a subtle but critical challenge for model evaluation. A model can learn *syntactic shortcuts*: rather than learning the true input-to-target mapping, it learns to recognize that a sample belongs to a known group of samples in the same graph structure, and predicts the average property of that group — effectively cheating by exploiting group superficial similarities and associated properties instead of learning the deeper mapping of sample to target. This can be seen as a special case of overfitting. Consequently, if the test set contains samples from graph structures present in training, evaluation does not measure generalization but merely whether the model learned these shortcuts, causing metrics to become optimistically biased and misrepresentative of true real-world performance. A properly constructed evaluation set must therefore be built exclusively from samples whose graph structures are absent from training. This motivates the need for a principled data partitioning mechanism designed specifically for datasets arising from RGPs.

### 3.2 Algorithm Overview

In the case where the graph structure is latent, the relations between samples are unknown. We can infer those relations given an appropriate distance function $f$ where nearby samples in the graph are expected to be close:

$$f : X \times X \to [0, \infty). \tag{4}$$

Crucially, this does not require all samples in a same graph structure to be mutually close, but rather that local distance along the path is reflected in the distance metric. Formally, the expected distance between nearby samples in the graph is much smaller than the expected distance between arbitrary samples from the distribution $\mathcal{D}$:

$$\mathbb{E}_{g\sim\mathcal{G}}\mathbb{E}_{(x,x')\sim\mathcal{D}_g\times\mathcal{D}_g}[f(x,x')|\delta(x,x')\leq k] \ll \mathbb{E}_{(x,x')\sim\mathcal{D}\times\mathcal{D}}[f(x,x')], \tag{5}$$

where $\delta(\cdot,\cdot)$ corresponds to the distance between two sample in the graph structure.

The inferred relations are then naturally modelled as a proximity graph $G$, where each sample is a node $v \in V$ and an edge is added between two nodes whenever their distance falls below a proximity threshold $\tau$:

$$G = (V, E), \quad E = \{(v_i, v_j) \mid v_i, v_j \in V, i < j, f(v_i, v_j) < \tau\}. \tag{6}$$

The threshold $\tau$ encodes the maximal accepted distance to plausibly consider two samples related. Thus, connected subgraphs in the proximity graph can be interpreted as the inferred graph structure from the generative process, see Figure 1 bottom.

The dataset partition is then performed at the connected component level. Formally, the components $C_1, ..., C_k$ form a partition of the dataset:

$$C = \{C_1, ..., C_k\}, \quad C_i \cap C_j = \varnothing \quad \forall i \neq j, \quad \bigcup_{i=1}^{k} C_i = D, \tag{7}$$

and we assign each component index entirely to either training $I_{\text{train}}$ or evaluation $I_{\text{test}}$:

$$I_{\text{train}} \cap I_{\text{test}} = \varnothing, \quad I_{\text{train}} \cup I_{\text{test}} = \{1, ..., k\}, \tag{8}$$

ensuring that no component is ever split across sets:

$$D_{\text{train}} = \bigcup_{i\in I_{\text{train}}} C_i, \quad D_{\text{test}} = \bigcup_{i\in I_{\text{test}}} C_i. \tag{9}$$

This guarantees that all evaluation samples originate from graph structures $g$ absent from training, provided the connected components correctly recover the latent graph structures.

### 3.3 HNSW Approximation

Constructing the proximity graph naively requires computing all pairwise distances, resulting in $O(n^2)$ complexity. Here, we leverage the Hierarchical Navigable Small World (HNSW) algorithm [24] to bypass this computation entirely and directly construct the proximity graph in $O(n \log(n))$ complexity. HNSW makes minimal assumptions about the data, and the distance function, operating correctly even for metrics that do not satisfy the triangle inequality, making it applicable to similarity-based functions as well.

We introduce two modifications to the standard algorithm. First, the standard HNSW caps the number of neighbors each node can hold, causing nodes with many true neighbors — hub nodes — to miss edges and degrading the accuracy of the resulting proximity graph. To address this, we maintain an explicit list of all edges computed during the build process and rely on this edge list for graph construction, recovering otherwise missed edges without significantly impacting runtime. Second, the standard HNSW algorithm maintains a candidate neighbor list of fixed size *ef*, evicting the worst candidate when full. We modify this rule: if the candidate being evicted has a distance below $\tau$, it is retained and the list is allowed to grow. This ensures no true neighbor is discarded during graph traversal at build time, improving recall.

### 3.4 Graph Community Detection

With the proximity graph constructed efficiently, we turn to the question of how to partition it. While theoretically grounded, the connected component formulation can degenerate on some densely connected datasets, where a single component may encompass the majority of nodes — or in the worst case, the entire dataset. This severely limits the degrees of freedom available when constructing the split, potentially making a balanced train/test split impossible.

To relax this constraint, we can optionally subdivide connected components into smaller, densely connected clusters called communities in graph terminology [25], increasing the granularity of the partition. We leverage the Leiden algorithm [23] for this purpose, which offers an effective community detection approach. The Figure 2a illustrates the comparison between connected component clustering and community clustering.

### 3.5 Post-Filtering

Splitting along communities rather than connected components introduces a trade-off: training samples may retain edges to test samples, allowing residual information leakage across the split boundary. As an optional post-processing step, we introduce post-filtering: all training samples sharing at least one edge with a test sample are removed. While this reduces the amount of available training data, it eliminates potential leakage. Furthermore, as community boundaries are by definition sparsely connected externally, the number of removed samples is small Figure 2b.

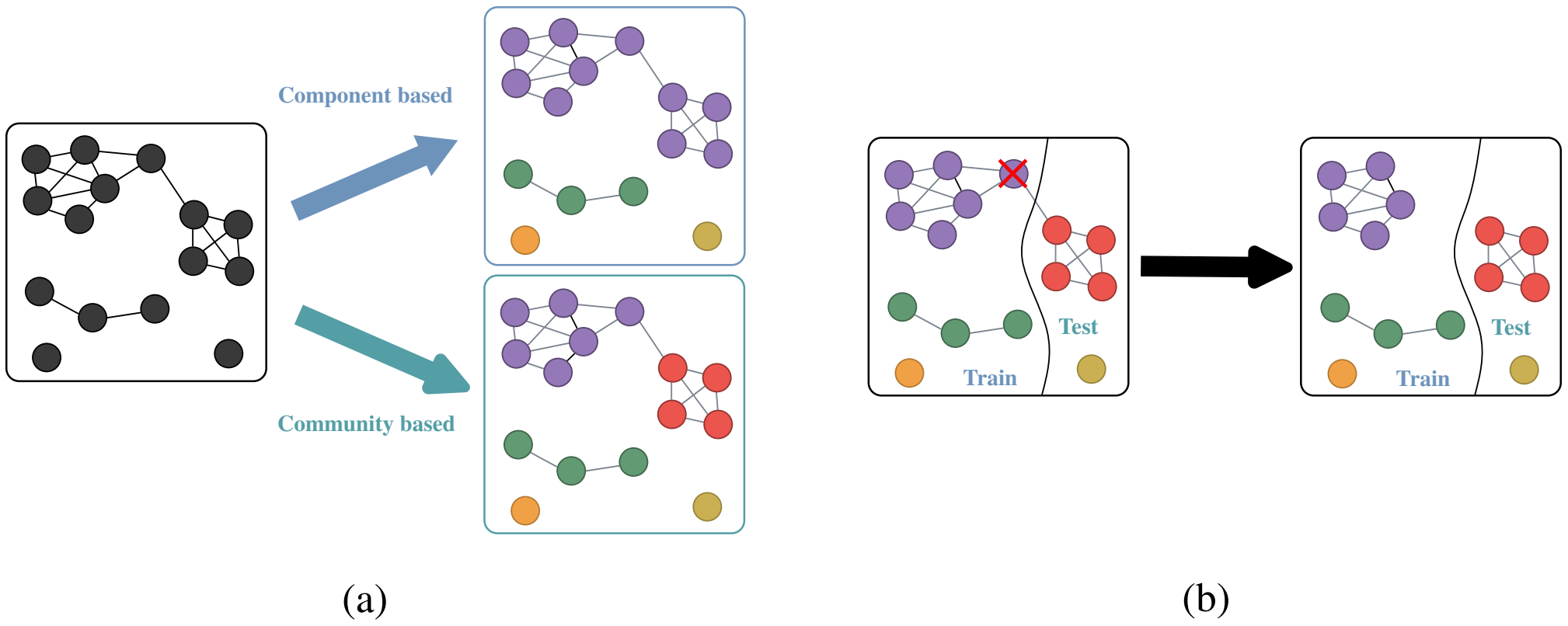

(a) (b)

Figure 2: The proposed splitting procedure comprises two stages: graph-based cluster construction and optional post-filtering to eliminate residual connections between the training and test sets. (a) Comparison of component-based and community-based clustering on the same proximity graph. In the component-based split, connected components are used directly as clusters. In the community-based split, the Leiden algorithm subdivides dense regions of each component into smaller communities, increasing the granularity of the partition and providing more degrees of freedom for the train/test split. Colors denote independent clusters. (b) Illustration of the optional post-filtering step. The curved line separates training samples (left) from test samples (right). Training samples sharing an edge with a test sample, marked with a red cross, are removed, eliminating residual information leakage at the partition boundary.

### 3.6 Fast and User Friendly

We implement Refnd in Rust for near-zero overhead performance, with Python bindings installable via pip for accessibility. The package exposes tools for proximity graph construction, community detection, nearest neighbor search, dataset partitioning, and dataset statistics. For each computationally intensive operation, both an exact $O(n^2)$ and an approximate $O(n \log(n))$ are provided, allowing users to select the appropriate version for their use case.

Refnd currently ships with three distance functions, referred to as *kernels* in our framework terminology: sequence identity for proteins, peptides and nucleotides via the Parasail C library [26], Template Modelling score for protein, DNA and RNA structures [27] and Tanimoto similarity for small molecules. Additional kernels will be added in future releases, and users can define custom kernels to support other data types.

## 4 Experimental Setup

We empirically validate Refnd on an antimicrobial peptide regression task derived from the curated Database of Antimicrobial Activity and Structure of Peptides (DBAASP) [16] through the Quantitative Mapping of Antimicrobial Peptides (QMAP) benchmark package [15]. Peptides are short chains of amino acids, and antimicrobial peptides inhibit the growth of or kill microorganisms. We retain only peptides with a measured minimum inhibitory concentration (MIC), a potency measurement against *Escherichia coli*, a standard measure of antimicrobial potency in which lower values correspond to higher activity, and use the corresponding peptide sequence as input. Because MIC values span several orders of magnitude, we regress the target in log-transformed space and predict $\log_{10}(\text{MIC})$. After filtering, the dataset contains 11,059 peptide sequences. This setting is particularly suitable for validating split quality because the dataset contains many closely related sequences and is therefore highly susceptible to information leakage.

Under the RGP interpretation, peptides sampled from the same graph structure form a peptide family. The evaluation objective is therefore family-level generalization: test peptides should come from families absent from training.

To isolate the effect of the split strategy from the effect of model capacity, we use fixed representations and two lightweight downstream predictors. Each peptide is embedded with the pretrained ESM-C 300M protein language model [28], used as a frozen feature extractor. For each sequence, residue-level embeddings are mean pooled across residue tokens, excluding beginning-of-sequence, end-of-sequence, and padding tokens, to obtain a single 960-dimensional representation. On top of these fixed embeddings, we train two predictor heads: a linear regression head and a two-layer multilayer perceptron (MLP) with 1,920 hidden units and ReLU activation. This design intentionally keeps the predictive model simple so that observed differences can be attributed to the split protocol rather than to architectural complexity.

We compare three split strategies. The first is a traditional random split applied at the sample level. The second is a cluster-based split where clusters are built with MMseqs2 [8] using `easy-cluster` with a 50% sequence identity threshold and the following parameters: `--alignment-mode 3`, `--gap-open 11`, `--gap-extend 1`, `--cov-mode 1`, `--cluster-mode 1`, `--comp-bias-corr 0`, and `-s 7.5`. Sequence identity is a similarity function widely used for protein data [29] [30]. Whole clusters are then assigned to either train or test. The third is the proposed Refnd split with community detection and post-filtering. We construct a proximity graph using the global protein sequence kernel at a 50% identity threshold. For graph construction, we use the library defaults of HNSWState, namely `m=16`, `m_max=16`, `m_max0=32`, `m_l=0.36`, `ef_init=1`, and `ef_construction=128`, together with `post_filtering=true`. Communities are detected with the default Leiden parameters `gamma=1.0`, `beta=0.01`, and `n_iterations=10`. For all three strategies, the split operation is applied twice: first to construct the held-out test set, and then again on the remaining training portion to obtain a validation set.

All downstream training settings are held constant across split strategies. Both models are optimized with Adam using a learning rate of $10^{-3}$ and a batch size of 64. The linear model is trained for 100 epochs without early stopping. The MLP is trained for up to 300 epochs with early stopping based on validation loss and a patience of 15 epochs. Mean squared error is used as the training objective.

We repeat each split-and-train experiment over random seeds 1 through 10 and report the mean and standard deviation across runs. Our primary evaluation metric is the Pearson Correlation Coefficient (PCC) on the test set, where higher values indicate better performance, defined as

$$\mathrm{PCC}(y, \hat{y}) = \frac{\sum_{i=1}^{n} (y_i - \overline{y})\left(\hat{y}_i - \overline{\hat{y}}\right)}{\sqrt{\sum_{i=1}^{n} (y_i - \overline{y})^2}\sqrt{\sum_{i=1}^{n} \left(\hat{y}_i - \overline{\hat{y}}\right)^2}}. \tag{10}$$

In addition to predictive performance, we evaluate split quality by measuring, for each test sequence, the sequence identity to its nearest neighbor in the training set. We refer to this quantity as max-identity. This auxiliary analysis is critical for evaluating the information leakage permitted by a split.

Table 1: Protocol summary of the evaluated split strategies. Only the split changes across experiments; embeddings, models, optimization, and evaluation remain fixed.

| Split | Construction | Threshold | Notes |
|---|---|---|---|
| Random | Sample-level random partition | None | Strong family leakage expected |
| MMseqs2 | Cluster first, then assign full clusters to train/test | 50% identity | Heuristic clustering can miss similar peptide pairs |
| Refnd | HNSW proximity graph + community detection + post-filtering | 50% identity | Heuristic graph construction from true similarity |

Because Refnd removes samples through the post-filtering step, its effective train, validation, and test sizes are smaller than those of the random and MMseqs2 splits. To control for this confound, we make size matching part of the experimental protocol: for each seed, after constructing the Refnd split, we subsample the random and MMseqs2 train set to match the corresponding effective Refnd sizes for that same seed. The matched sizes vary across seeds because the Refnd partition itself varies with the sampled split. In the saved runs used here, the matched sizes average 5,193 training samples, 1,435.6 validation samples, and 2,240.5 test samples.

## 5 Results

We evaluate the three split strategies along two complementary axes: the degree of information leakage permitted by each split, measured by the max-identity distribution of test sequences against the training set, illustrated in Figure 3, and observed evaluation performances, summarized in Table 3. We present the leakage analysis first, as it directly explains the performance differences observed across splits.

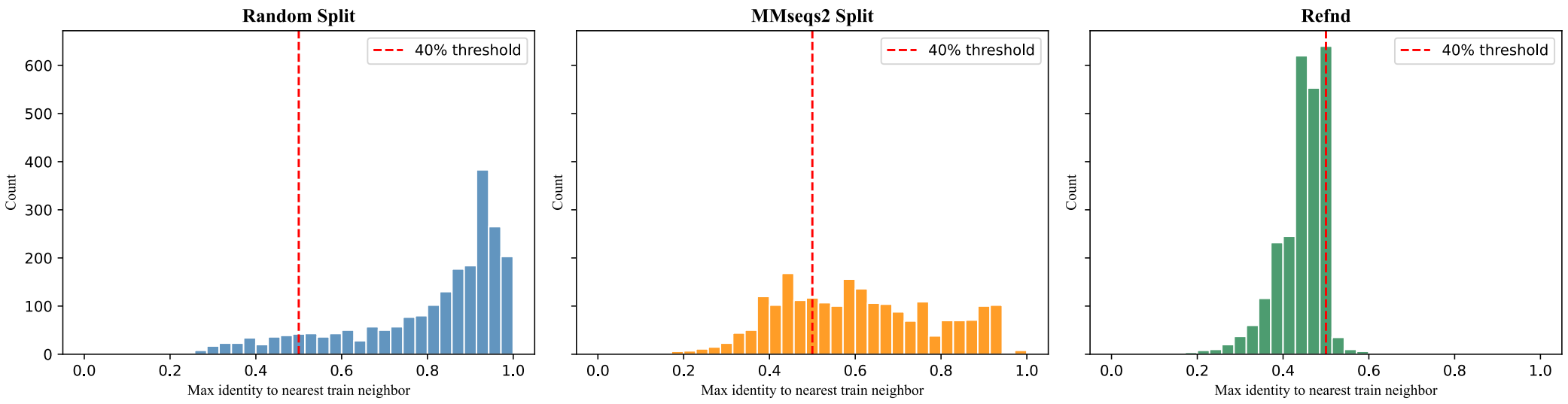

Figure 3: Distribution of max-identity, the maximum sequence identity between each test peptide and its nearest training neighbor, for the three split strategies. The dashed red line marks the 50% identity threshold used to define family-level separation for MMSeqs2 and Refnd.

### 5.1 Identity Distributions Reveal How Much Leakage Each Split Permits

The max-identity distributions in Figure 3 directly characterize the degree of information leakage permitted by each split protocol. Under the random split, a large proportion of test sequences have nearest training neighbors well above the 50% identity threshold, with many extreme cases close to or equal to 100%. MMseqs2 shifts the distribution downward but still leaves a substantial fraction of violations. In contrast, the Refnd split concentrates the distribution below the threshold, leaving only a thin tail of residual violations. These rare violations are consistent with the approximate nature of the HNSW proximity graph used by Refnd.

### 5.2 Predictive Performance Reflects Leakage Severity

The random split yields the highest average PCC across both models — 0.599 for the linear model and 0.664 for the MLP — yet substantially overestimates generalization performance because it allows strong sequence overlap between train and test, as shown in Figure 3. In this regime, a model can achieve strong test performance by interpolating within peptide families already represented in the training set, rather than by learning transferable rules that generalizes to unseen families. This data leakage is what explains the strong PCC performances.

The effect of leakage becomes even clearer when a randomly split test set is stratified by nearest-neighbor identity to the training set, as shown in Table 2. On the subset of test sequences whose nearest training neighbor has identity at most 50%, the PCC drops to 0.448 for the linear model and 0.498 for the MLP. In contrast, on the subset whose nearest training neighbor has identity at least 80%, the PCC rises to 0.631 and 0.729, respectively. The high-identity subset contains 1,494 samples, compared with only 231 samples in the low-identity subset, meaning the random split test set is overwhelmingly composed of family-overlapping sequences. The aggregate performance reported under a random split therefore reflects predominantly shortcut learning rather than genuine generalization, and substantially overestimates the model's ability to predict antimicrobial potency for novel peptide families.

Table 2: Leakage analysis under a representative random split. Performance is substantially higher on test peptides that have highly similar training neighbors.

| **Random-split subset** | **Samples** | **Linear PCC** | **MLP PCC** |
|---|---|---|---|
| Nearest train neighbor $\leq 50\%$ identity | 231 | 0.448 | 0.498 |
| Nearest train neighbor $\geq 80\%$ identity | 1,494 | 0.631 | 0.729 |

MMseqs2 [8] provides a meaningful improvement over random splitting, with PCC dropping to 0.553 for the linear model and 0.575 for the MLP. However, as shown in Figure 3, it does not fully enforce the intended 50% sequence-identity boundary on this peptide dataset, which could result in residual information leakage. Notably, the gap between the linear model and the MLP narrows substantially — from 0.065 under the random split to 0.022 — suggesting that part of the MLP advantage was driven by information leakage rather than genuine model capacity. The MMseqs2 split provides a better separation than random splitting, but it does not fully enforce the intended threshold, leaving the possibility of residual information leakage.

Table 3: Test **Pearson Correlation Coefficient (PCC)** across split strategies, reported as mean ± standard deviation over 10 different dataset splits with different random seeds. The training pipeline and parameters are kept the same, only the split is modified. All pairwise differences between splits are statistically significant ($p < 0.01$, Wilcoxon signed rank) — see supplementary materials.

| **Model** | **Random** | **MMseqs2** | **Refnd** |
|---|---|---|---|
| Linear | 0.599 ± 0.018 | 0.553 ± 0.035 | 0.516 ± 0.039 |
| MLP | 0.664 ± 0.019 | 0.575 ± 0.031 | 0.505 ± 0.036 |

Refnd provides the strongest enforcement of the intended 50% sequence-identity boundary. As shown in Figure 3, Refnd strongly limits the number of train-test pairs whose sequence identity exceeds the threshold, yielding a much stricter family-level separation than the alternative splits. Predictive performance under this split drops substantially for both models, with both reaching roughly 0.51 PCC on average. The large MLP advantage observed under the random split also disappears, and the linear model slightly outperforms the MLP on average under this protocol. Refnd produces the most conservative evaluation, and likely reflects the most faithful estimate of real-world generalization performance.

## 6 Conclusion

Our experiments demonstrate that Refnd yields lower but likely more realistic evaluation performance than both random and MMSeqs2-based splits, since model architecture is kept identical across experiments. The observed performance drop is attributable to a reduction of information leakage rather than a reduction of model performance. This is further supported by the stratified analysis, in which models evaluated on test samples close to the training set (>80% identity) achieve 77% PCC, compared to 42% PCC on distant samples (<50% identity), directly demonstrating that proximity to training data drives observed performance and that random splits yield over-optimistic estimates. These results motivate the adoption of Refnd for any dataset arising from a RGP.

The primary parameter requiring tuning is the proximity threshold $\tau$, which encodes the expected distance between inference data and training data, making the split interpretable and application-aware. To determine an appropriate value, we recommend the methodology introduced by QMAP [15]: sample data representative of the production distribution, measure the distance from each sample to its nearest neighbor in the training set, and set $\tau$ such that the test-to-train distance distribution approximates the inferred production-to-train distance distribution. This grounds the choice of $\tau$ in the intended deployment context rather than an arbitrary heuristic.

While validated on antimicrobial peptides, the RGP formulation predicts broad applicability to any dataset containing relations between samples that can be attributed to dependency graph structures. Pharmaceutical small molecules, where compounds share common scaffolds with targeted modifications optimizing efficacy, constitute a direct instantiation of the RGP. Similarly, protein structures exhibit reused structural domains and patterns across homologs, and nucleotide sequences undergo the same incremental mutational process as protein sequences. We note, however, that the RGP assumption does not hold universally across biochemical datasets; researchers should therefore carefully consider the generative process underlying their data before applying Refnd.

Beyond dataset splitting, the proximity graph constructed by Refnd encodes relational structure that can be leveraged for broader applications. At the dataset level, the connectivity structure of the graph provides natural statistics to characterize dataset diversity and redundancy, directly useful for dataset curation and quality assessment in biochemical machine learning. At the retrieval level, the HNSW approximation enables $O(\log(n))$ nearest neighbor search with high recall under arbitrary distance or similarity metrics, extending the utility of Refnd to large-scale database search beyond the dataset partitioning setting.

## 7 Data & Code Availability

Refnd Package: https://github.com/anthol42/refnd
Experimental Code: https://github.com/JacobCote/refnd_experimental_code
Dataset: https://github.com/anthol42/QMAP

## 8 Acknowledgments

Lavertu A. acknowledges funding from the Fonds de recherche du Québec through a Master's Research Scholarships (doi: 10.69777/369917) and NSERC's Canada Graduate Research Scholarship – Master's program. Côté J. acknowledges funding from the Fonds de recherche du Québec through a Master's Research Scholarships (doi: 10.69777/360097) and NSERC's Canada Graduate Research Scholarship – Master's prosgram. Corbeil J. acknowledges the support of the Canadian Biomanufacturing Research Fund (CBRF) and, specifically, the PandemicStopAI project: an accelerated response to bacterial pandemics. He also acknowledged the funds for the CQDM (Quebec Consortium for Drug Discovery). Gobeil S. is supported by an NSERC Discovery Grant (RGPIN-2023-03546), an NSERC Discovery Launch Supplement (DGECR-2023-00091) and a CBRF and Biosciences Research Infrastructure Fund (BRIF) award for the "Biologics RAMP-UP: Biologics Rapid Actuation of Mass Production Under Pandemic conditions" project (CBRF2-2023-00176).

# A Supplementary Material

## A.1 Additional Formal Definitions

The following equations provide additional formal grounding for the definitions introduced in the main text.

### A.1.1 Distance Function Properties

The distance function $f : X \times X \rightarrow [0, \infty)$, Equation (4) is subject to the identity of indiscernibles:

$$f(x, y) = 0 \Leftrightarrow x = y \quad \forall x, y \in X \tag{11}$$

and symmetry:

$$f(x, y) = f(y, x) \quad \forall x, y \in X \tag{12}$$

but, importantly, not the triangular inequality.

## A.2 Statistical Significance of Split Comparisons

To verify that the performance differences between split strategies are statistically significant, we apply the Wilcoxon signed-rank test to the per-seed PCC scores across all pairwise comparisons. This nonparametric paired test is appropriate here for two reasons: (1) with only 10 seeds, normality of the differences cannot be reliably verified, making a parametric t-test unsuitable; and (2) the seeds provide a natural pairing structure, as each seed controls for model initialization variance across all three splits simultaneously.

Table 4: Pairwise Wilcoxon signed-rank test results comparing test PCC across split strategies over 10 seeds. All differences are significant at $p < 0.01$. The uniform p-value of 0.002 is the minimum achievable with $n = 10$ paired observations, indicating that the ordering between splits was consistent across every seed without exception.

| Comparison | Model | Δ mean PCC | p-value |
|---|---|---|---|
| Random vs MMseqs2 | Linear | −0.049 | 0.002 |
| Random vs MMseqs2 | MLP | −0.097 | 0.002 |
| Random vs Refnd | Linear | −0.103 | 0.002 |
| Random vs Refnd | MLP | −0.176 | 0.002 |
| MMseqs2 vs Refnd | Linear | −0.053 | 0.002 |
| MMseqs2 vs Refnd | MLP | −0.078 | 0.002 |

## A.3 Dataset

The curated DBAASP dataset provided by QMAP counts 11,059 peptide sequence with lengths ranging from 2 to 190 residues with a median of 17 residues.